\documentstyle[prb,multicol,aps,epsf]{revtex}

\begin{document}
\title{Phase transitions in the antiferroelectric-ferroelectric mixed systems}
\author{V.A.Stephanovich*, M.D.Glinchuk**, L.Jastrabik***}
\address{*Institute of Physics of Semiconductors, NASc of\\
Ukraine,\\
Pr.Nauki 45, 252650 Kiev, Ukraine \\
**Institute for\\
Problems of Materials Science, NASc of Ukraine,\\
Krjijanovskogo\\
3, 252180 Kiev, Ukraine \\
***Institute of Physics, Academy of\\
Sciences of the Czech Republic,\\
Na Slovance 2, 180 40 Praha 8,\\
Czech Republic }
\maketitle

\begin{abstract}
Phase diagram of solid solutions of antiferroelectric and ferroelectric
materials like PbZr$_{1-x}$Ti$_x$O$_3$ (PZT), Rb$_x$(NH$_4$)$_{1-x}$H$_2$PO$%
_4$ (RADP) is calculated at the region $x\leq 0.1$. Antiferroelectric and
ferroelectric order parameters (which correspond respectively to the
difference and sum of two sublattices ions shifts) were introduced for
antiferroelectric host lattice. Small admixture of ferroelectric component
was considered as a random electric field source. The influence of this
field on both aforementioned order parameters was calculated. Self
consistent averaging with calculated random field distribution function made
it possible to obtain the system of equations for order parameters of mixed
system. The antiferroelectric $T_a$ and ferroelectric $T_c$ phase transition
temperature concentrational dependence was expressed through the transition
temperatures of the composition and members. The obtained $T_a(x)$
dependence was shown to fit pretty good the observed values for PZT and
RADP. The application of the proposed theoretical approach to other types of
mixed systems is discussed.
\end{abstract}
\begin{multicols}{2} \narrowtext
\section{Introduction}

Mixed antiferroelectric-ferroelectric systems like PbZr$_{1-x}$Ti$_x$O$_3$
(PZT), Rb$_{1-x}$(NH$_4$)$_x$H$_2$PO$_4$ (RADP) ($0\leq x\leq 1$) attract
much attention of scientists because of their unusual properties and phase
diagram peculiarities. These anomalies made PZT material useful for many
technical application [1]. In particular compositions of PZT enriched by Zr (%
$x<0,1$) appeared to be useful for application in the novel branches of
electronic technique (see e.g. [2], [3] and ref. therein). Theoretical
description of the properties and phase diagram nearby antiferroelectric
phase transition is hard problem because of complexity of description of
antiferroelectric phase itself. The proximity of free energies of
antiferroelectric and ferroelectric state is known to be the characteristic
feature of antiferroelectrics [4]. This proximity lead to dielectric
permittivity anomalies and to the possibility to induce ferroelectric order
by application of small enough electric field. The latter can be the clue of
the reasons of ferroelectric phase appearance even at small $x$ values.
Small content of ferroelectric material in the mixed system can be
considered as the source of electric dipoles and induced by them random
electric field. Calculation of this field magnitude allowing for
dipole-dipole interaction via ferroelectric and nonferroelectric soft modes
had shown that mean field is induced by the interaction via ferroelectric
soft mode whereas other soft modes increase mean square field [5]. The
supposition that nonzero internal mean field can destroy antiferroelectric
order by the same way as external field made it possible to describe
semiquantitively the observed phase diagram [5].

In present work we performed detailed calculations of phase diagram of mixed
antiferroelectric-ferroelectric system for the region of small concentration
of ferroelectric material. The obtained data fitted pretty good measured
phase diagram for PZT and RADP. The application of proposed method for
calculation of phase diagrams in different mixed ferroelectrics is discussed.

\section{General equations}

Let us begin with the description of antiferroelectric phase. In Ising model
free energy $F$ can be written as [4]:

\begin{eqnarray}
&&F=\frac{1}{2}J_{S}\sigma ^{2}+\frac{1}{2}J_{A}S^{2}-p^{\ast
}E\sigma- \nonumber \\
&&-\frac{T%
}{2}\ln \left[ 2ch2\beta (p^{\ast }E+J_{S}\sigma )+2ch2\beta J_{A}S\right]
\label{e1}
\end{eqnarray}
where $\beta \equiv 1/(kT)$, effective dipole moment $p^{\ast }=(\varepsilon
_{a}-1)\gamma _{a}p/3$, $\varepsilon _{a}$ and $\gamma _{a}$ are
antiferroelectric component susceptibility and Lorentz factor respectively, $%
p=Zb$ ($\pm b$ are the positions of two-well potential minima), $E$ is
electric field, $\sigma $ and $S$ are respectively ferroelectric and
antiferroelectric transition order parameters ($\sigma =\overline{\xi }/b$, $%
S=\eta /b$, $\overline{\xi }$ and $\eta $ being mean homogeneous and
inhomogeneous displacements respectively); $\ J_{A}=T_{A}$ and $J_{S}=T_{S}$ 
are respectively antiferroelectric transition and Curie
temperatures. Note, that in this model mean dipole moment of the unit cell
has the form:

\begin{equation}
d_r=Z(\overline{\xi }\pm \eta )  \label{e2}
\end{equation}

Thus we consider the simplest two-sublattice model of antiferroelectric.

\subsection{Order parameters}

Eq. (\ref{e1}) leads to the following expressions for the order parameters:

\begin{eqnarray}
\sigma &=&\frac{sh2\beta (p^{*}E+J_S\sigma )}{ch2\beta (p^{*}E+J_S\sigma
)+ch(2\beta J_AS)}  \label{e3} \\
S&=&\frac{sh2\beta J_AS}{ch2\beta (p^{*}E+J_S\sigma )+ch(2\beta J_AS)}
\nonumber
\end{eqnarray}

Eqs. (\ref{e1})-(\ref{e3}) describe antiferroelectric component, i.e. mixed
system at $x=0$. Ferroelectric component corresponds to $x=1$ and its order
parameter $L$ in Ising model has the form [4]

\begin{equation}
L=th\beta (E_0L+d^{*}E)  \label{e4}
\end{equation}
where $E_0=T_{cmf}$ is mean field, that defined ferroelectric transition
temperature, $d^{*}$ is effective dipole moment of ferroelectric component
ions.

Addition of small amount of ferroelectric component to antiferroelectric one
leads to appearance of random electric field $\varepsilon $, induced by
electric dipoles of this component in the mixed system. Mean field in the
mixed system has to include the contribution from ferroelectric component
and from ferroelectric order parameter of antiferroelectric component namely

\begin{equation}
\overline{\varepsilon }=J_S(1-x)\sigma +xE_0L  \label{e5}
\end{equation}

This field as the most probable one has to define the position of maximum of
random field distribution function $f(\varepsilon ,\sigma ,L)$. It is equal
to $\delta (\varepsilon -J_S(1-x)\sigma -xE_0L)$ in mean field
approximation. Beyond of this approximation one has to take into account
random field, i.e. to substitute $E+\varepsilon $ ($E\neq 0$) or $%
\varepsilon $ ($E=0$) for $E$ in Eqs. (\ref{e3}), (\ref{e4}) and to perform
averaging with random field distribution function. This yields:

\[
\sigma =\int\limits_{-\infty }^\infty \frac{sh(2\beta p^{*}(\varepsilon
+E))f(\varepsilon ,\sigma ,L)d\varepsilon }{ch(2\beta p^{*}(\varepsilon
+E))+ch(2\beta J_A(1-x)S)}
\]

\begin{equation}
S=\int\limits_{-\infty }^\infty \frac{sh(2\beta J_A(1-x)S)f(\varepsilon
,\sigma ,L)d\varepsilon }{ch(2\beta p^{*}(\varepsilon +E))+ch(2\beta
J_A(1-x)S)}  \label{e6}
\end{equation}

\[
L=\int\limits_{-\infty }^\infty th(\beta d^{*}(\varepsilon +E))f(\varepsilon
,\sigma ,L)d\varepsilon
\]

One can see that in mean field approximation when $f(\varepsilon ,\sigma ,L)$
has the form of $\delta $-function Eqs. (\ref{e6}) give Eqs. (\ref{e3}) at $%
x=0$ and Eq. (\ref{e4}) at $x=1$.

Eqs. (\ref{e6}) are general expressions for order parameters dependence on
concentrations and characteristics of mixed system components. Note, that in
the case $S=0$ (both components of mixed system are ferroelectrics) the
equations for $\sigma $ and $L$ have the same form as it has to be expected.

\subsection{Transition temperatures}

Let us proceed now to calculations of the transition temperatures $T_a$ (to
antiferroelectric phase) and $T_c$ (to ferroelectric phase) dependence on
concentration. For the phase transitions of the second order these
temperatures can be obtained from Eqs. (\ref{e6}) in the limit of $\sigma
\rightarrow 0$, $S\rightarrow 0$, $L\rightarrow 0$. In such a limit one can
simplify the integrands in Eqs. (\ref{e6}) substituting $ch2\beta \
J_AS\approx 1$, $sh2\beta \ J_AS\approx 2\beta \ J_AS$ and representing the
distribution function in the form:

\begin{equation}
f(\varepsilon ,\sigma ,L)=f_0(\varepsilon )-\left( \frac{J_S(1-x)\sigma }{%
p^{*}}+\frac{E_0xL}{d^{*}}\right) \left( \frac{df}{d\varepsilon }\right)
_{\sigma =L=0}  \label{e7}
\end{equation}
Here $f_0(\varepsilon )$ is the distribution function in the case of mean
field absence.

Keeping in mind that $f_0(\varepsilon )$ is even parity function in
aforementioned limit Eqs. (\ref{e6}) give:

\[
\sigma =\left[ (1-x)J_S\sigma +xE_0L\frac{p^{*}}{d^{*}}\right] \beta I(x)
\]

\begin{equation}
S=(1-x)\beta J_ASI(x)  \label{e8}
\end{equation}

\[
L=\left[ (1-x)J_S\sigma \frac{d^{*}}{p^{*}}+xE_0L\right] \beta I(x)
\]

\begin{equation}
I(x)=\int\limits_{-\infty }^\infty \frac{f_0(y)dy}{ch^2(\beta y)}  \label{e9}
\end{equation}
Integral (\ref{e9}) has to depend on concentration because of dependence on $%
x$ of the distribution function width.

The solution of the system of Eqs. (\ref{e8}) yields:
\begin{mathletters}
\begin{equation}
T_a=(1-x)J_AI_a(x),T_c=\left[ (1-x)J_S+xE_0\right] I_c(x)  \label{e10a}
\end{equation}
or in dimensionless variables

\begin{equation}
\tau _a=\frac{T_a}{T_A};\tau _c=\frac{T_c}{T_S};\lambda =\frac{T_{cmf}}{T_S}
\label{e10b}
\end{equation}
Eq. (\ref{e10a}) can be rewritten as

\begin{equation}
\tau _a=(1-x)I_a(x);\tau _c=\left[ x\lambda +(1-x)\right] I_c(x)
\label{e10c}
\end{equation}
Here $I_a(x)$ and $I_c(x)$ are integrals (\ref{e9}) at $\beta =\beta _a$ and
$\beta =\beta _c$ respectively. One can see, that the transition
temperatures are defined by the corresponding transition temperatures and
concentrations of the components of mixed system and random field
distribution function characteristics.

Equations (10) define concentrational dependence of transition temperature,
i.e. phase diagram of mixed system for small concentration of ferroelectric
component. One can see, that Eqs. (10) give the correct values in mean field
approximation:

$T_a=T_A$, $T_c=T_S$ at $x=0$ and $T_a=0$, $T_c=E_0=T_{cmf}$ at $x=1$.

\subsection{Critical concentrations}

Random field induced by electric dipoles of ferroelectric component of mixed
system can destroy completely antiferroelectric phase at some concentration
of the dipoles. This concentration $x=x_{ca}$ is known to be critical
concentration at which $T_{a}=0$, so that antiferroelectric phase exists at $%
x<x_{ca}$ only. On the other hand ferroelectric phase in the mixed system
arises at $x>x_{cc}$ where $x_{cc}$ is critical concentration for
ferroelectric phase transition. It has to be $T_{c}\geq T_{A}$ at $x\geq
x_{cc}$. Therefore critical concentrations $x_{ca}$ and $x_{cc}$ in the
region $x\ll 1$ can be obtained from the equations $T_{a}=0$ ($x_{ca}$) and $%
T_{c}=T_{A}$ ($x=x_{cc}$). Let us limit ourselves by the case $\Delta =0,$
which is of \ importance both for case of \ PZT and RADP on the ''edges'' of
phase diagram. To perform the calculations on the base of Eqs. (\ref{e10c})
we have to know concentrational dependence of \ integrals $I_{a,c}(x)$
defined by the distribution function $f_{0}(y)$. For electric dipoles as
random field sources it can be represented in Gaussian form (see [6] for
details):

\end{mathletters}
\begin{equation}
f_0(y)=\frac 1{2\sqrt{\pi c}}\exp \left( -\frac{y^2}{4c}\right)  \label{e11}
\end{equation}

\begin{equation}
c=x\frac{16\pi }{15}\frac{d^{\ast 2}}{\varepsilon _{0}^{2}a^{3}r_{c}^{3}}\xi
\equiv c_{0}x  \label{e12}
\end{equation}

Here $\varepsilon _{0}$, $a$ and $r_{c}$ are respectively dielectric
permittivity, lattice constant and correlation radius connected with soft
ferroelectric mode of antiferroelectric component of mixed system. Parameter
$\xi \geq 1$ takes into account dipole-dipole interaction via
antiferroelectric and other nonferroelectric soft modes [5]. Performing
changing of variables in integrals $I_{a,c}$ ($\beta mE=u$), we obtain that
this reduces just to renormalization of parameter $c_{0}$ in (\ref{e12}),
giving $c_{m}=c_{0}m^{2},\ m=p^{\ast },d^{\ast }$.

To map correctly the interval $0<z<\infty $ into interval $0<x<1$ we adopt
following parametrization for $z_{c,a}$%
\begin{eqnarray}
&&z_{c,a}=\varphi _{c,a}(x)\rho ^{3},\ \rho =\frac{r_{c}}{a},\ \varphi _{c}(x)=%
\frac{1-x}{1-(1-x)^{\mu }},\nonumber \\ 
&&\varphi _{a}(x)=\frac{x}{1-x^{\mu }},\ \mu
\rightarrow \infty .  \label{e12a}
\end{eqnarray}
Let us begin with critical concentration for antiferroelectric phase
transition.

At $\beta =\beta _{a}$ Eqs.(\ref{e10a}) with respect to Eqs. (\ref{e11}), (%
\ref{e12}) at $\beta _{a}$ $\rightarrow \infty $ yields

\begin{equation}
I_{a}(x)=\frac{1}{\sqrt{x}}\frac{1}{\sqrt{\pi }}\frac{\tau _{a}}{q_{a}}%
;q_{a}=\frac{c_{p^{\ast }}}{T_{A}}  \label{e13}
\end{equation}

Substitution of Eq. (\ref{e13}) into the expression for $\tau _a$ (see Eqs. (%
\ref{e10c})) gives the following equation for critical concentration $x_{ca}$%
:

\begin{equation}
\frac{x_{ca}}{1-x_{ca}^{\mu }}=\frac{1}{\pi q_{a}^{2}}.  \label{e14}
\end{equation}
Eq. (\ref{e14}) has been solved numerically during the calculation of $\tau
_{a}.$

Critical concentration for ferroelectric phase transition can be obtained by
the same way, but in the limit of $T_{c}\rightarrow T_{A}$, i.e. at $\tau
_{c}\rightarrow T_{A}/T_{S}$, this ratio being larger or close to unity.
Since in such a limit integral $I_{c}(x)$ can be calculated only numerically
we shall obtain $x_{cc}$ value in the next section devoted to numerical
calculations with two fitting parameters $\lambda =T_{cmf}/T_{S}$ and $%
q_{c}=c_{d^{\ast }}/T_{S}$. 

\section{Phase diagram of mixed antiferro-ferroelectric system}

Phase diagram. i.e. concentrational dependence of antiferroelectric and
ferroelectric phase transition temeprature was calculated with help of
Eqs.(10). Integrals $I_{c,a}(x)$ were calculated numerically. The obtained
phase diagram is represented in Fig.1 in the region of small concentration
of ferroelectric component for several values of dimensionless parameters $%
q_a,$ $q_c$ and $\lambda $ (see Eq.(\ref{e10b})). One can see from Fig.1
that antiferroelectric phase transition critical concentration $x_{ca}$
increases with $q_a$ decrease, its values being close to those calculated on
the base of Eq.(\ref{e14}). The obtained critical concentration of
ferroelectric phase transition several times smaller than that for
antiferroelectric transition. This seems to be the result of proximity of
free energies and may be of $T_S$ and $T_A$ which is characteristic feature
of antiferroelectrics. The dimensionless transition temperature to
antiferroelectric phase $\tau _a$ decreases with concentration increase for
all the considered parameters, whereas the behaviour of that to
ferroelectric phase $\tau _c$ depends on the choise of the parameters
values. In particular $\tau _c$ can increase or decrease with increase of
electric dipoles concentration (see Fig.1). Physically these types of
behaviour depend on distribution function width and ratio ($\lambda $) of
ferroelectric phase transition temepratures $T_{cmf}$ of ferroelectric
component of mixed system to that of antiferroelectric component $T_S$: the
smaller the distribution function width and the larger $\lambda $ value, the
larger ferroelectric phase transition temperature in the mixed system.
\section{Discussion}

4.1. Let us begin with comparison of obtained data with experiment. The
prominent example of mixed ferro-antiferro system is known to be PbZr$_{1-x}$%
Ti$_{x}$O$_{3}$ ($0\leq x\leq 1$), where PbZrO$_{3}$ is antiferroelectric
with transition temeprature from cubic paraelectric phase to rhombic
antiferroelectric phase at $T_{A}=500$ K and PbTiO$_{3}$ is ferroelectric
with transition temeprature from cubic paraelectric phase to tetragonal
ferroelectric phase at $T_{cmf}=763$ K. Phase diagram of mixed PZT system
was investigated in many papers, but mainly at $T\geq T_{room}$ (see e.g.
[6] and ref. therein). Therefore critical concentration $x_{ca}$ is known,
and $T_{a}=300$ K at $x=(5-6)\%$. One can see that the curve for $q_{a}=1.7$
gives $T_{a}\simeq 300$ K at $x=0.06$ and $x_{ca}\approx 0.11$.  General
form of calculated $\tau _{a}$ concentrational dependence looks like
experimental curve at $x<0.1$ ( Fig.1). There is a scattering of
experimental data for ferroelectric phase transition temperature at small
PbTiO$_{3}$ content ($x<0.05$) because of influence of some uncontrolled
defects and impurities on $T_{c}$ value. The most probable behaviour of $%
T_{c}$ is its slight increase (Fig.1) [6]. Such behaviour is similar to the
calculated curve with $q_{c}=1.7$ and $\lambda =7$.

Another important example of mixed ferro-antiferro system is Rb$_{1-x}$(NH$%
_{4}$)$_{x}$H$_{2}$PO$_{4}$ ($0\leq x\leq 1$) as a representative of
hydrogen-bonded material. It was studied in many details (see e.g. [7]). RbH$%
_{2}$PO$_{4}$ (RDP) undergoes a ferroelectric phase transition at $%
T_{cmf}=146$ K, NH$_{4}$H$_{2}$PO$_{4}$ (ADP) becomes
antiferroelectric near $T_{A}=148$ K. In mixed RADP system
antiferroelectric transition temperature $T_{a}$ decreases with
increase of RDP content and $T_{a}=0$ K near $x=0.2$ of RDP
admixture, i.e. $x_{ca}\approx 0.2$ [8]. Substitution of this
$x_{ca}$ value into Eq.(\ref{e14}) gives $q_{a}\approx 1$. One can
see from Fig.2 that the curve, which corresponds to $q_{a}=1.175$
fits the experimental points. Contrary to PZT there is no data
about existence of ferroelectric phase transition at small
($<10\%$) content of RDP. The reason for this difference can be
the approximately equal values of $T_{cmf}$ and $T_{A}$ in RADP
whereas they are different in PZT. Because of this theoretical
parameter $\lambda \approx T_{A}/T_{S}\sim 1$  in RADP and at
$\lambda =1$ $\tau _{c}=T_{c}/T_{S}$ curve may be lower than $\tau
_{a}$ curve and only antiferroelectric phase transition exists.
One can see from Fig.2, that proposed theory explaines the
observed behaviour of ferroelectric phase transition decrease with
increase of antiferroelectric component concentration and
observed critical concentration $x_{cc}$ ($q_{c}=1.424,\ \lambda
=1.1$). Note that tuneling has to be taken into account for mixed
systems of KDP family.
Preliminary calculations have shown, that tuneling result into
change of fitting parameters values.

Therefore the proposed theory describes pretty good phase diagrams of PZT
and RADP for small $x$ values. Note, that only for small $x$ ($x<0.2$) Ti
and Rb ions can be considered as impurities randomly distributed in PbZrO$%
_{3}$ and NH$_{4}$H$_{2}$PO$_{4}$ matrices so that they can be the sources
of random field.

4.2. The proposed theory can be applied for another types of mixed systems.
For example, mixture of two ferroelectric systems $(F_{1})_{1-x}(F_{2})_{x}$
at $x<0,1$ described with the help of Eqs. (\ref{e6}), (\ref{e8}) at zero
antiferroelectric phase order parameter ($S=0$). In this case the Equations
for $\sigma $ and $L$ completely the same, i.e. $\sigma \equiv L$. In such
situation mean field $E_{0}=T_{cmf}$ is ferroelectric phase transition
temperature of $F_{1}$ component whereas $F_{2}$ component defines the form
and width of random field distribution function. Because the main physical
idea of the proposed model is the inhibition and distruction of long range
ferroelectric or antiferroelectric order by random electric field of
different sources this model can be useful for description to relaxor
ferroelectrics like PbMg$_{1/3}$Nb$_{2/3}$O$_{3}$ (PMN), PbSc$_{1/2}$Ta$%
_{1/2}$O$_{3}$ (PST), Pb$_{1-y}$La$_{y}$Zr$_{0,65}$Ti$_{0,35}$O$_{3}$ (PLZT,
$y=0,08$; 0,09). But the relaxors like PMN, PST are not solid solutions of
two component, and in PLZT La ions are the impurities. Thus in all these
materials there is onlu one ferroelectric component known as Burns reference
phase (see [9] and ref. therein), which is PbZr$_{0,65}$Ti$_{0,35}$O$_{3}$
in the of PLZT. The transition temperature from para- to ferroelectric phase
of the reference phase $T_{cmf}\equiv E_{0}$ defines the position of random
field distribution maxima whereas the substitutional disorder, vacancies of
lead and oxygen, impurities like La in PLZT define the distribution function
width. Namely such a model has been proposed recently [10]. It permitted to
describe the observed phase diagram of the relaxors, their dynamic and
static properties peculiarities (see e.g. [11] and ref. therein).

Since the model is able to describe the distraction of long range order by
any random field sources it can be useful also for description of solid
solutions of ferroelectric and dielectric components, e.g. Ba(Ti$_{1-x}$Sn$%
_{x}$)O$_{3}$ ($x<0,1$). The observed in this material relaxor-like
behaviour as well as change of phase transition order [12] can be the
consequence of strong random field influence as it was shown recently
[11,13]. Note that some residual ferroelectric domains were observed in
another relaxor PLZT (9/65/35) [14]. As the matter of fact, the dipole glass
and mixed ferroglass phases (which appear in the relaxors with temperature
lowering) can be considered as the reentrant phases with respect to this
residual ferroelectric phase existing near the Burns temperature.

Therefore the generality of proposed model gives a possibility to apply it
to various mixed systems which are solid solutions of ferro- and
antiferroelectrics, two ferroelectrics, ferroelectric and dielectric etc. as
well as to ferroelectric relaxors.

\end{multicols}
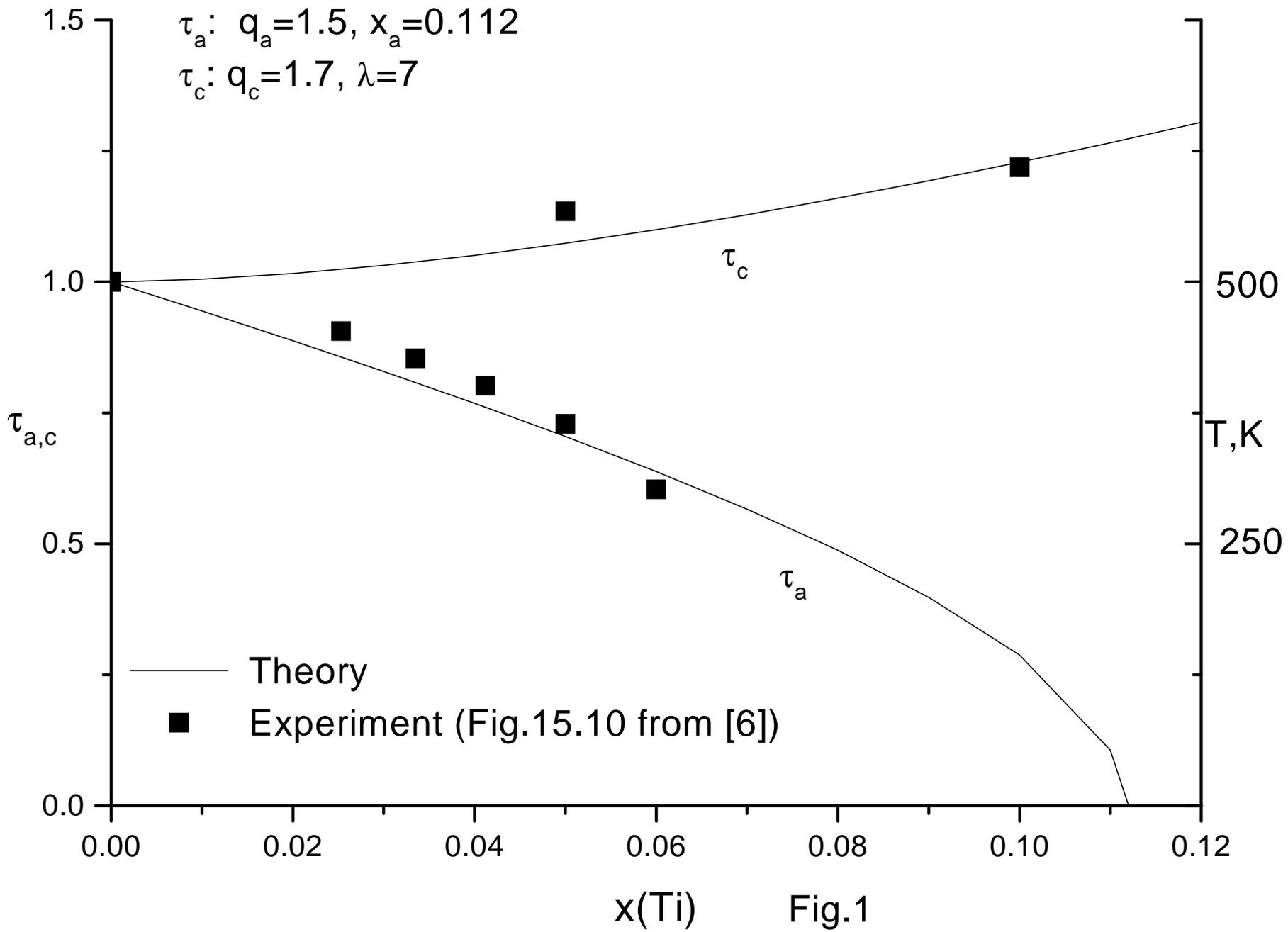
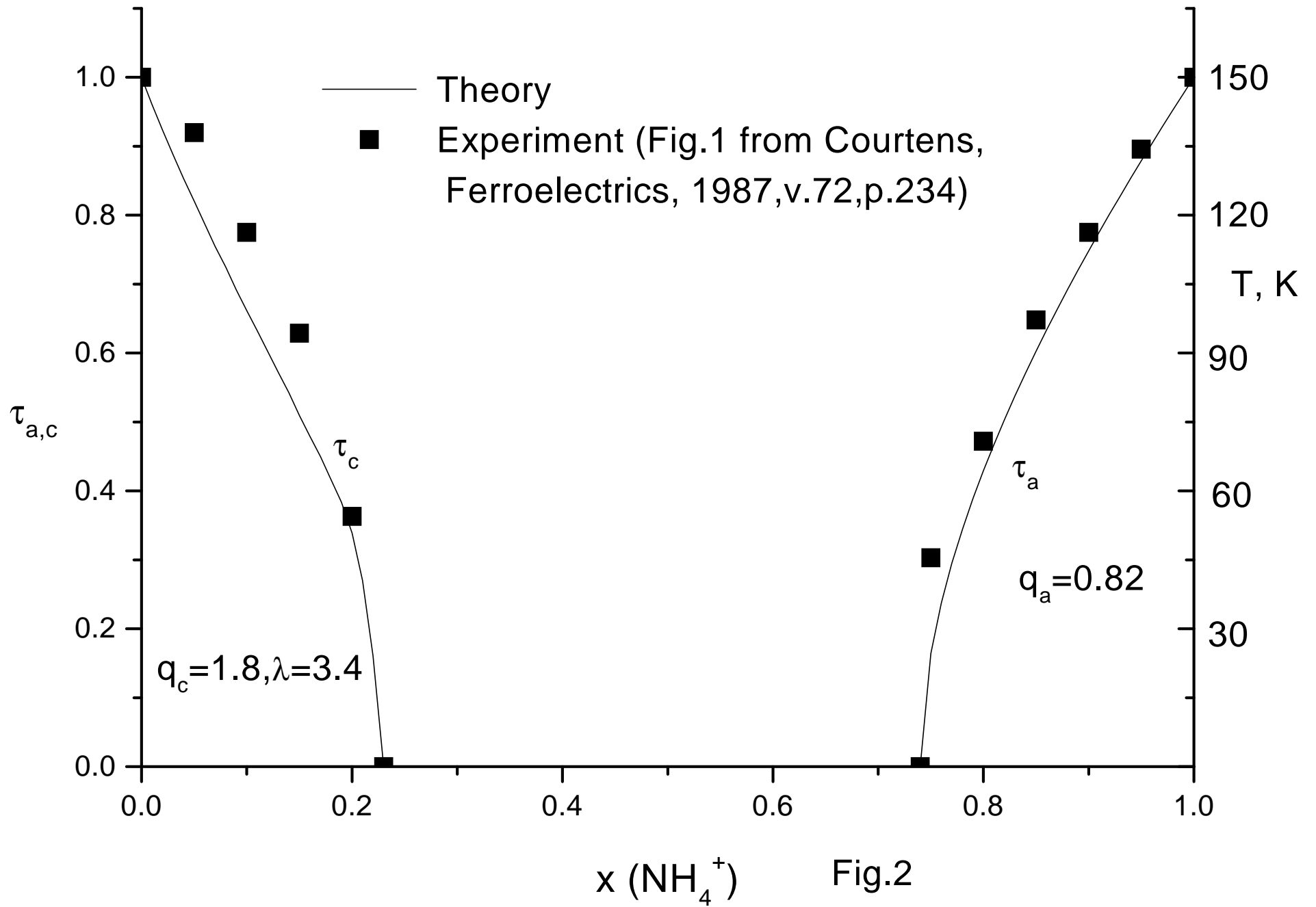
\end{document}